\documentclass[runningheads]{llncs}

\usepackage{eccv}
\usepackage{eccvabbrv}
\usepackage{graphicx}
\usepackage{booktabs}
\usepackage{algorithm}
\usepackage{subfloat}
\usepackage{algpseudocode}
\usepackage{multirow}
\usepackage{enumitem}
\usepackage{wrapfig}
\usepackage{tcolorbox}

\usepackage[accsupp]{axessibility}  

\usepackage{hyperref}
\usepackage{xspace}

\usepackage{orcidlink}

\newcommand \method {\texttt{EditShield}\xspace}  

\begin{document}

\title{\texttt{EditShield}: Protecting Unauthorized Image Editing by Instruction-guided Diffusion Models} 

\titlerunning{EditShield}

\author{
  Ruoxi Chen\inst{1}\orcidlink{0000-0003-2626-5448} \and 
  Haibo Jin\inst{2}\orcidlink{0000-0002-7244-7659} \and 
  Yixin Liu\inst{3}\orcidlink{0000-0003-3856-439X} \and
  Jinyin Chen\inst{1}\thanks{Corresponding author} \and 
  Haohan Wang\inst{2}\orcidlink{0000-0002-1826-4069} \and 
  Lichao Sun\inst{3}
}

\authorrunning{R.~Chen et al.}

\institute{
  Zhejiang University of Technology, Hangzhou, China\\
\email{\{2112003149,chenjinyin\}@zjut.edu.cn}\\ \and
  University of Illinois Urbana-Champaign, Champaign, USA\\
  \email{\{haibo,haohanw\}@illinois.edu}
  \and
  Lehigh University, Bethlehem, USA\\
  \email{\{yila22,lis221\}@lehigh.edu}
}

\maketitle

\begin{abstract}
Text-to-image diffusion models have emerged as an evolutionary for producing creative content in image synthesis. Based on the impressive generation abilities of these models, instruction-guided diffusion models can edit images with simple instructions and input images. While they empower users to obtain their desired edited images with ease, they have raised concerns about unauthorized image manipulation. Prior research has delved into the unauthorized use of personalized diffusion models; however, this problem of instruction-guided diffusion models remains largely unexplored.
In this paper, we first propose a protection method \method against unauthorized modifications from such models. Specifically, \method works by adding imperceptible perturbations that can shift the latent representation used in the diffusion process, tricking models into generating unrealistic images with mismatched subjects. 
Our extensive experiments demonstrate \method's effectiveness among synthetic and real-world datasets. 
Besides, we found that \method performs robustly against various manipulation settings across editing types and synonymous instruction phrases. 
\end{abstract}

\section{Introduction}
\label{sec:intro}

In the domain of image generation, the emergence of diffusion models~\cite{sohl2015deep,ho2020denoising,saharia2022photorealistic} has marked a new era of innovation, especially in text-to-image generation tasks. These large-scale models
stand out for their ability to produce photorealistic and artistic images. Notably, Stable Diffusion (SD)~\cite{rombach2022high} and DALL-E 3~\cite{betker2023improving} are capable of generating compelling images from simple text prompts in seconds. Their widespread adoption fuels millions of applications in areas ranging from digital art to content creation.

These text-to-image diffusion models have also emerged as powerful tools for image editing~\cite{avrahami2022blended,wang2023imagen,kawar2023imagic,sheynin2023emu}. Instruction-driven image editing models~\cite{brooks2023instructpix2pix,zhang2023magicbrush,fu2023guiding,huang2023smartedit} represent a significant advancement, can modify the given source image according to the instructions specified by the user. Remarkably, these advancements facilitate modifications directly during the forward pass, eliminating the need for additional inputs such as reference images, segmentation masks, or model fine-tuning. This flexibility also benefits practicality as such guidance is more aligned with human intuition.

The success of the instruction-guided image editing also presents avenues for misuse with serious ethical implications. The malicious editor may create misleading narratives or ``fake news'', thereby contributing to public misinformation \cite{Misinformation1, Misinformation2}. Similarly, unauthorized modifications of celebrity images or personal photographs—ranging from altering facial expressions to creating inappropriate content - pose grave concerns for privacy, consent, and reputational harm. Artists may see their works manipulated without permission \cite{totonews}, leading to misunderstandings, economic losses, and a diminution of artistic integrity~\cite{thedarkside,rawimpact}. 


\begin{figure}[t]
\centering
\includegraphics[width=0.85\linewidth]{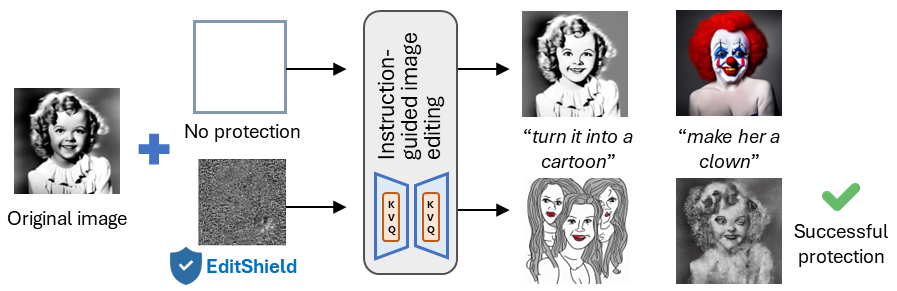}
\caption{The illustration of protection by \method. Editors cannot get their expected images with the protection of \method.}
\label{example_scenario}
\end{figure} 


Several attempts~\cite{salman2023raising,ShanCW0HZ23,van2023anti,LiangWHZXSXMG23} have been proposed to address the unauthorized modifications by diffusion models and have shown effectiveness through experiments. These solutions, however, focus on preventing personalized editing, which requires reference images and the fine-tuning process. The protection benefits from fixed and pre-defined text prompts used in personalization, facilitating the development of specific countermeasures. In contrast, instruction-guided image editing, characterized by its dynamic and user-specific instructions, presents a significant challenge to existing protective approaches. This distinction underscores the urgent need for innovative research focused on developing robust protections for instruction-guided editing, regardless of editing types and instruction phrases.


To tackle these challenges, in this paper, we investigate how to protect images against unauthorized editing from instruction-guided diffusion models, as depicted in Fig.~\ref{example_scenario}. We design \method, a protection method designed to safeguard images from being unauthorized edited by instruction-guided diffusion models. 
It operates by introducing strategic perturbations that effectively disrupt the latent representations these models rely on, leading to significant mismatches between the intended editing outcomes and the actual results, often manifesting as compromised image quality and mismatched subjects. Through extensive experiments on both synthetic and real-world datasets, we verify the effectiveness of \method across diverse images and is robust against various manipulation settings under different editing types and synonymous instruction. 

Our contributions are summarized as follows:
\begin{itemize} [nolistsep, leftmargin=*]
    \item We take the first step to study the potential risk of unauthorized image editing from instruction-guided diffusion models. We then develop \method, to safeguard images from unauthorized modifications by these models.
    \item EditShield works by introducing small perturbations that can disrupt the latent representation leveraged by those models, resulting in unrealistic images with mismatched subjects after editing. 
    \item Through extensive experiments, we show that our method effectively protects a wide range of images from unauthorized editing and exhibits robustness against diverse editing types and instruction phrases. 
\end{itemize}

\section{Background}
\subsection{Diffusion Models} 
Motivated by the non-equilibrium statistical physics, diffusion models~\cite{sohl2015deep,ho2020denoising,nichol2022glide,ramesh2022hierarchical,saharia2022photorealistic,gu2022vector} have shown superior performance in text-to-image generation tasks, notably excelling in producing images of remarkable quality and diversity. Typically, they employ a pre-trained model such as CLIP~\cite{radford2021learning} or T5~\cite{raffel2020exploring} for extracting text embeddings, which are then leveraged by the generation component for synthesizing images. 

These models distinguish themselves from language models by utilizing continuous latent representations grounded in the Markov Chain framework. Given an input image $x_0 \sim q(x)$, the forward diffusion process adds a small amount of Gaussian noise to the sample in $T$ step that produces a sequence of noisy samples $\{x_1, x_2, ...,x_T\}$. The step sizes are controlled by a variance schedule $\{ \beta_t \in (0,1)\}_{t=1}^T \}$. We can sample $x_t$ at any arbitrary time step $t$, as follows:
\begin{equation}
    x_t = \sqrt{\bar \alpha_t}x_0+\sqrt{1-\bar \alpha_t}\epsilon
\end{equation}
where $\alpha_t=1-\beta_t$, $\bar \alpha_t=\prod_{i=1}^t \alpha_i$ and $\epsilon \sim \mathcal{N} (0,\mathbf{I})$. Conversely, the backward process uses the reverse Markov Chain to denoise the sample $x_{t}$ from $x_{t+1}$. The injected noise in each step can be learned by the denoising model $\epsilon_\theta$, which is trained to minimize the distance between the estimated noise and the true noise.

With the integration of a Variational Auto-Encoder (VAE)~\cite{kingma2013auto} trained on images, latent diffusion models (LDMs)~\cite{rombach2022high,podell2023sdxl} significantly enhance the efficiency and quality of the generated images while concurrently reducing computational costs. They are now widely employed in visual generation~\cite{blattmann2023align,avrahami2023blended,blattmann2023stable} and editing~\cite{corneanu2024latentpaint}.
Given an image $x$, the image encoder $\mathcal{E}$ first encodes it into a latent representation $z=\mathcal{E}(x)$. Then, the noise-adding and denoising operations are conducted on encoded latent $z$. 
The unconditioned latent diffusion models are trained:
\begin{equation}
     \mathcal{L}_{unc}=\mathbb E_{\mathcal{E}(x), \epsilon \sim \mathcal{N} (0,1),t }\left[||\epsilon-\epsilon_\theta (z_t,t)||_2^2 \right]
\end{equation}
where $\mathcal{L}_{unc}$ denotes the loss function of unconditional diffusion models, $t$ is the time step, and $z_t$ is the latent noise at time step $t$.

\subsection{Instruction-guided Image Editing}
Text-guided image editing ~\cite{avrahami2022blended,couairon2022diffedit,wang2023imagen,kawar2023imagic,sheynin2023emu,chen2024subject} have demonstrated the efficacy of leveraging pre-trained diffusion models, in the nuanced realm of image editing such as subject-specific modifications and style transfers. Besides, some models \cite{GalAAPBCC23,ruiz2023dreambooth} enable users to achieve highly customized edits, through fine-tuning with some reference images and text prompts.

Different from them, instruction-guided image editing accepts straight commands like ``add fireworks to the sky'' for guidance, without the need for additional descriptions or regional masks. Brooks et al.~\cite{brooks2023instructpix2pix} first proposed InstructPix2Pix based on Prompt-to-Prompt~\cite{hertz2022prompt}. The model leverages texts generated by a fine-tuned version of GPT-3~\cite{brown2020language} and images generated by SD. Subsequent works are all built upon InstructPix2Pix and further enhanced the performance by fine-tuning it with the MagicBrush dataset~\cite{zhang2023magicbrush}, incorporating human feedback~\cite{zhang2023hive} for better alignment, and with the assistance from multi-modal large language models~\cite{fu2023guiding,huang2023smartedit}. 
These models are based on latent diffusion and introduce input images and instruction prompts for guidance. 
Given the image conditioning $c_I$ and text conditioning $c_T$, the network is trained to predict the noise added on the latent, for image content preserving and instruction following. The objective of latent diffusion is minimized as follows:
\begin{equation}
    \mathcal{L}_{ins}=\mathbb E_{\mathcal{E}(x),\mathcal{E}(c_I), c_T, \epsilon \sim \mathcal{N} (0,1),t }\left[||\epsilon-\epsilon_\theta (z_t,t,\mathcal{E}(c_I),c_T)||_2^2 \right]
\end{equation}
where $\mathcal{L}_{ins}$ denotes the loss function of instructional image editing models.

\begin{figure}[t]
\centering
\includegraphics[width=0.75\linewidth]{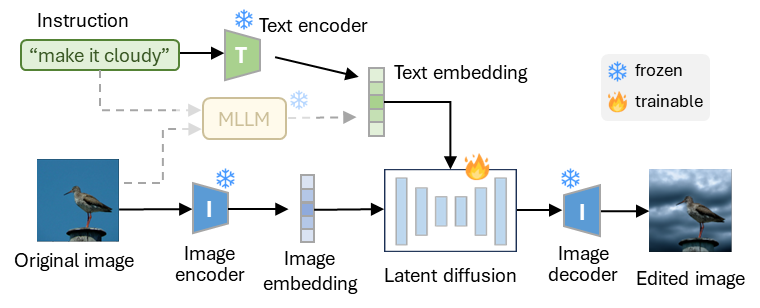}
\vspace{-5pt}
\caption{The workflow of instructional image editing based on diffusion models.}
\label{ip2p-bkg}
\end{figure}

As Fig.~\ref{ip2p-bkg} suggests, the user inputs a source image and an instruction prompt to the model, where the image and text are encoded by the image and text encoder or MLLMs, respectively. The latent diffusion model is trained under the guidance of both image embedding and text embedding. 
Finally, the edited image will be reconstructed by the image decoder from its latent representation.

In this paper, we mainly target InstructPix2Pix~\cite{brooks2023instructpix2pix}model, and assess the generalization of our method to its enhanced fine-tuned version on MagicBrush~\cite{zhang2023magicbrush}, since these models are open-sourced and publicly available until now.


\subsection{Protection Against Unauthorized Use}
As diffusion models rapidly evolve, concerns regarding its misuse cannot be overlooked.
To address this issue, various strategies have been developed to safeguard public images of individuals against diffusion models. The main idea involves introducing specially crafted noise to the image. For instance, PhotoGuard~\cite{salman2023raising} introduces imperceptible perturbations to images, effectively raising the computational cost for malicious editing attempts. As a result, LDMs cannot generate realistic
images when attempting to edit upon the perturbed image. In the field of artworks and paintings, Shan et al.~\cite{ShanCW0HZ23} developed Glaze which can protect artists against style mimicry by text-to-image diffusion models. Similarly, Liang et al.~\cite{LiangWHZXSXMG23} adopted the concept of adversarial examples for protecting images from being learned, imitated, and copied by diffusion models. Van Le et al.~\cite{van2023anti} generated subtle noise to the image published by the user, to disrupt the generation quality of DreamBooth trained on these perturbed images. 
These protective measures target copyright infringers who utilize diffusion models with specific prompts tied to particular subjects. However, their effectiveness is challenged in real-world scenarios where models encounter a variety of unseen prompts, potentially making these protections ineffective. Furthermore, the specific problem posed by instruction-guided diffusion models—where user-generated instructions dictate the content creation process - remains unexplored by existing works.


\section{Preliminary}

\subsection{Problem Statement}
We first introduce the detailed protection setting and assumptions, where we consider a two-player game: the Image Editor for instructing diffusion model and the Image Protector for the protection method. 

\begin{figure}[htbp]
\centering
\vspace{-10pt}
\includegraphics[width=0.7\linewidth]{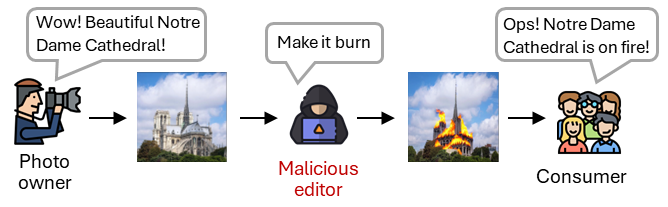}
\caption{The image editing scenario that we consider. }
\label{scenario}
\end{figure}

\noindent \textbf{Protection Scenario.} Instruction-guided editing models can be convenient for modifying images. But they may also be exploited by unauthorized use towards a specific individual or community. 
Within this context, we identify three key entities in the lifecycle of information generation, modification, and consumption: the photo owner, who creates or is depicted in the images; the malicious editor, who intends to modify the image for unauthorized use for malicious purposes, such as spreading fake news; and the consumer, who may unknowingly encounter and believe the manipulated content. This flow from content creation to its unauthorized modification and eventual consumption is depicted in Fig.~\ref{scenario}. To counteract these risks, the protector helps shield images from such manipulations, thereby making unauthorized edits ineffective.

\noindent \textbf{Image Editor.} The editor's goal is to modify the image from the image owners without permission. The image editor can access the well-trained, instruction-guided image editing models and then exploit them to modify images for potentially harmful purposes, such as negatively impacting one's appearance. Furthermore, the editor can obtain images from the social platform from their owners, bypassing any permissions or ethical considerations. The editing process is straightforward: the editor feeds the source image into the model and provides specific instruction prompts.

\noindent \textbf{Image Protector.} The goal is to add noise to the image to prevent unauthorized editing. The protector can use our method to generate perturbations and inject them into the image before publishing it online. The protection is successful if the image has a mismatched subject identity or is of awful quality after editing. The protector can be the image owner or another third-party service.

We assume the protector is aware that the editor will use instruction-guided diffusion models for image editing. The protector has white-box access to those editing models. This assumption is now common among works~\cite {salman2023raising,LiangWHZXSXMG23} since more models are open-sourced online. However, the protector does not know the instructions and editing type specified by the editor. Besides, we assume the protector has significant computational power for generating perturbations.

\subsection{Design Challenges}
Existing solutions~\cite{LiangWHZXSXMG23,van2023anti}, have showcased effectiveness in scenarios involving personalized image editing~\cite{GalAAPBCC23,ruiz2023dreambooth}, where protections are designed around specific prompts tied to given reference images and pre-defined subjects. These personalized image editing models operate under the premise of fixed text prompts, like ``a photo of sks [class noun]'', where sks specifies the subject and ``[class noun]'' is the subject category. However, these protection mechanisms face challenges in instruction-guided editing scenarios, where the image protector lacks access to the dynamic instruction prompts and editing types employed by unauthorized editors. The absence of specific prompts poses a substantial challenge in designing an effective protection strategy.

\section{Our Proposed Method} 
\subsection{Design Intuition}
With unknown instructions that unauthorized editors might use, designing protection perturbations may be difficult. Given that the input images are accessible to the protector, our solution explores the potential within the image space itself. 

Image editing aims to enact desired modifications through language commands while retaining as much of the original image's detail as possible. The unauthorized editing aims to modify images without losing inherent content and quality. For example, an instruction to \textit{``replace the dog with a cat''} should not accidentally transform the surrounding lawn into a desert. Such unintended changes, while not fulfilling the editor's intent, present an opportunity for protection by exploiting these mismatches. 

Inspired by this, we propose leveraging content changes as a protection mechanism against malicious editing, independent of the specific instructions. Our approach involves crafting protective noise designed to disturb the editing process guided by a spectrum of potential instructions.



We target the latent representation that is leveraged by the mainstream of instruction-guided diffusion models. We introduce a instruction-independent protection method \method aims to shift the latent representation of the original image while maintaining the protection noise imperceptible. After that, following the instruction-guided diffusion process, the edited image will be irrelevant or unrealistic, effectively safeguarding against unauthorized modifications.

\subsection{\method}
Based on our previous analysis, we derive the two conditions that the protected image should satisfy, then formulate \method, as an optimization problem based on those two conditions. 

\noindent \textbf{Latent inconsistency.} Recall that in instruction-guided image editing models, image conditioning and text conditioning are determined by the source image and the instruction prompt, respectively. So we can manipulate the source image to influence the image embedding. This can be achieved by shifting the latent representation before entering the latent diffusion process. In this way, the protection operates independently of the instruction prompt. Even if the editing follows the instructions provided by an unauthorized editor, it is not possible to obtain an edited image that maintains related subjects.

Note that the model first projects the source image into the image representation by VAE. To disrupt the image embedding in the latent space, we force the projected embedding of the protected image to deviate from the original representation. We define the latent inconsistency loss to quantitively measure the latent shift: $\mathcal{L}_{latent}=Dist((\mathcal{E}(x_p),\mathcal{E}(x))$, where $x_p$ and $x$ represent the protected and the source image, respectively. $\mathcal{E}$ is the VAE used in the instruction-guided editing model. $Dist(\cdot, \cdot)$ computes the distance between two latent representations. We use mean-square error (MSE) by default.

\noindent \textbf{Perceptual consistency.} While shifting latent representation, the protective noise should not hurt the visual quality of the source image. So, the protection should be human-invisible and should not affect image semantics. Visible noise could alert an unauthorized editor to the presence of protection measures, prompting them to either discard the image, conduct data transformations such as filtering, or apply specific high-resolution editing instructions to circumvent the protection. Therefore, the noise must be well-crafted to be invisible to humans, ensuring that the image retains its original appeal and message while still effectively disrupting unauthorized editing processes. 
We define the perceptual consistency loss as $\mathcal{L}_{perceptual}={||x_p-x||}_{2}^{2}$ with $l_2$ norm regularization.

\noindent \textbf{Transformation Robustness.} The shift in latent representation, while effective in preventing unauthorized image modifications, can introduce mismatches in content and unrealistic textures, compromising the edited image's integrity. However, perturbations crafted directly with the above two conditions are fragile to minor transformations. To address this, we incorporate the Expectation Over Transformation (EOT)~\cite{athalye2018synthesizing}, into our optimization strategy. Specifically, given $F$ as a distribution over a set of transformations $f$, we integrate the transformation robustness loss into $\mathcal{L}_{latent}$ and rewrite it as $\mathcal{L}_{latent}=Dist((\mathcal{E}(f(x_p)),\mathcal{E}(x))$, where $f(x_p)$ is the transformed image of $x_p$ after applying the transformation $f$. This integration ensures that our protective perturbations remain effective and consistent across a variety of potential editing scenarios, including those involving subtle image transformations.

Combining the above losses together, our final optimization problem is:
\begin{equation}
    \max_{x_p} \mathbb{E}_{f\sim F} [Dist((\mathcal{E}(f(x_p)),\mathcal{E}(x))]- \beta \cdot ||x_p-x||_{2}^{2}
\end{equation}
where $\beta$ is the hyper-parameter to balance the two losses. To guarantee that the perturbation is human-invisible, the protection noise $\delta$ is bounded by the overall perturbation budget. 
We use projected gradient descent to
solve the optimization problem efficiently. In practice, we initialize $x_p$ by $x$ plus an extra small standard Gaussian noise $\mathcal {N}(0, \mathbf{I})$. This iterative process persists until the protected image reaches convergence or when the maximum number of iterations $S$ is obtained. After applying the perturbation generated by \method, users can publish their images online.
It can be seen from the above that \method generates untargeted perturbations. We don't set a specific target before optimization to make it more general towards input images consisting of different contents.

\section{Evaluation}
\subsection{Experiment Setup}
\noindent \textbf{Models.} Our objective is to prevent image editing from being used by instruction-guided diffusion models without permission. We choose \textit{InstructPix2Pix (ip2p)}~\cite{brooks2023instructpix2pix}, which is the mainstream tool and basic backbone for instruction-guided image editing. We use the ip2p model based on SD v1.5, along with its enhanced model \textit{ip2p-mb} that is fine-tuned on MagicBrush dataset~\cite{zhang2023magicbrush}.

\noindent \textbf{Datasets.} We evaluate the effectiveness of our protection using two datasets: a filtered subset from the synthetic dataset IPr2Pr~\cite{brooks2023instructpix2pix} and the test split of MagicBrush. The protection is applied to the source images, and instructions provided in the dataset are used for subsequent image editing.

\noindent \textbf{Baselines.} We compare our method with PhotoGuard~\cite{salman2023raising}, which adds imperceptible perturbations to prevent image editing with text prompts. The parameters are configured following the settings reported in the original paper.

\noindent \textbf{Evaluation Metrics.} Following~\cite{brooks2023instructpix2pix}, we calculate two metrics to assess image editing, as discussed below. 
\begin{itemize} [nolistsep, leftmargin=*]
    \item CLIP image similarity: We first extract image embedding from both the source and edited images without and with protection. We then calculate the cosine similarity of these CLIP image embeddings. This metric measures how much the edited image agrees with the source image. 
    \item CLIP text-image direction similarity~\cite{gal2022stylegan}: We adopt LLaVA-1.5 13B~\cite{liu2023improvedllava} for generating descriptions of both the source and edited images without and with protection. These descriptions, combined with the editing instructions, are used to compute CLIP text-image direction similarity. This metric measures how much the change in text captions agrees with the change in the images.
    \end{itemize}
For both metrics, smaller values indicate worse image editing capabilities and, consequently, better image protection capability. 

\noindent \textbf{Implementation Details.} We apply protections on 2,000 source images from IPr2Pr and 1,000 images from MagicBrush. Subsequently, these images are edited by instruction-guided diffusion models, with the guidance of the corresponding instructions provided in each dataset. For $T$ in EOT, we choose transformations that the editor might use, including Gaussian kernel smoothing (kernel=5, sigma=1.5), image rotation with rotate angle=5$^\circ$, and center cropping. 
We set the overall perturbation budget as 4/255, $\beta=0.2$, and $S$=30. To mitigate randomness, we use the same seed when editing source and protected images. We defer further experimental details to supplementary materials. 

\begin{figure*}[t]
\centering
\includegraphics[width=0.9\linewidth]{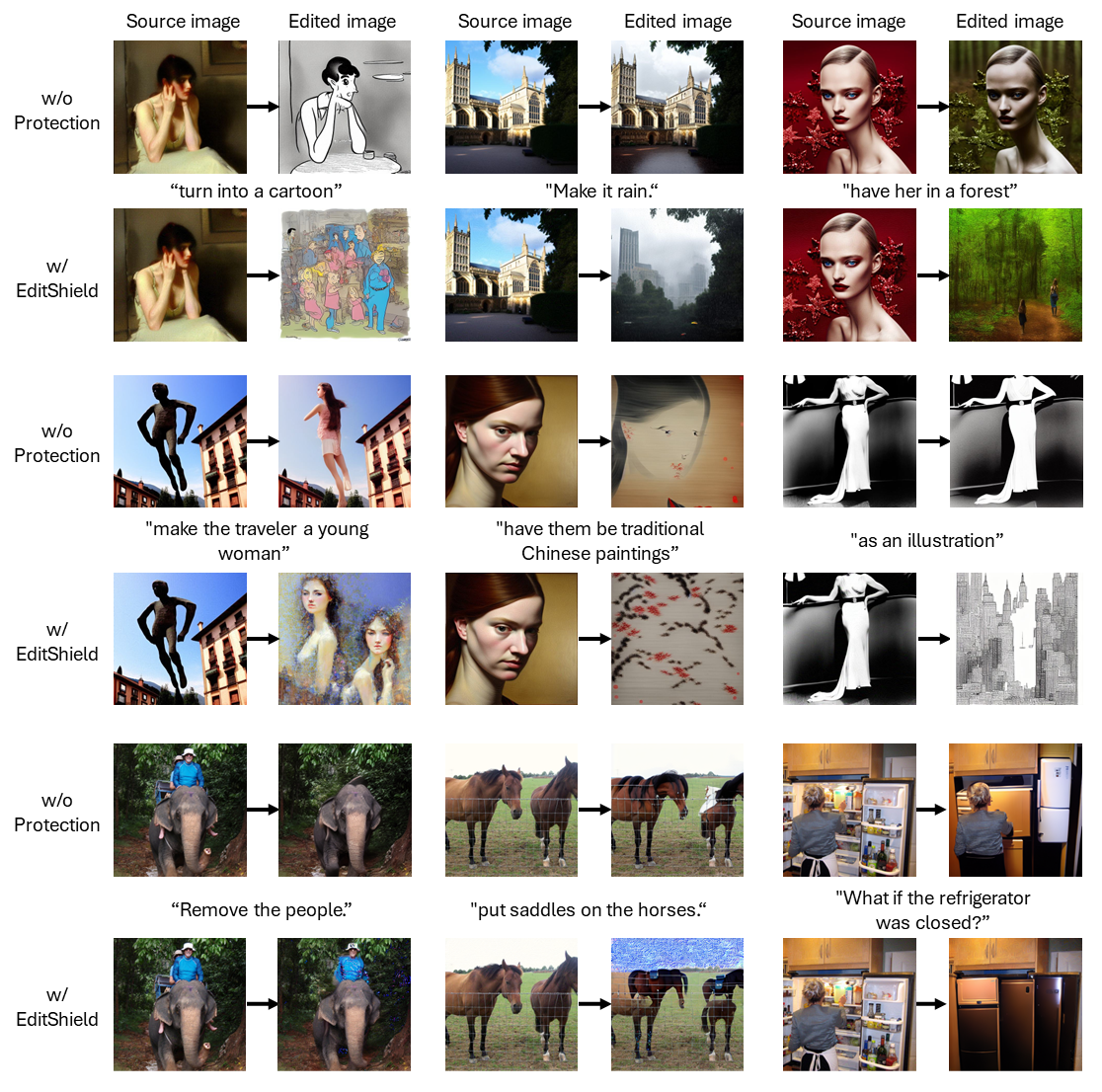}
\caption{Qualitative protection results. Each pair of rows displays the source and protected images with their respective edited versions. }
\label{VIS1}
\end{figure*}

\begin{figure*}[t]
\centering
\begin{minipage}[c]{0.68\linewidth}
\centering
 \subfloat[ip2p]{
        \includegraphics[width=0.49\linewidth]{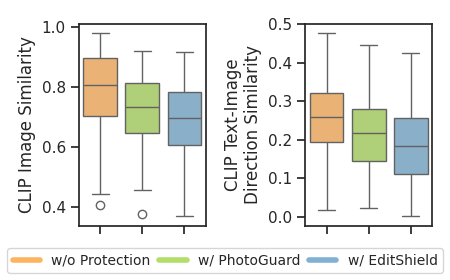}}
    \subfloat[ip2p-mb]{
        \includegraphics[width=0.48\linewidth]{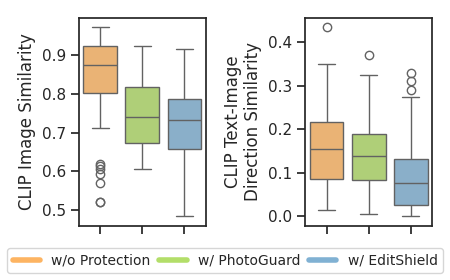}}
\caption{Quantitative results on editing with protection of PhotoGuard and \method. }
\label{tradeoff}
\end{minipage}%
\hfill
\begin{minipage}[c]{0.3\linewidth}
   \centering
   \huge
   \renewcommand\arraystretch{1.3}
   \resizebox{0.95\linewidth}{!}{
    \begin{tabular}{cccc}

        \toprule
        \textbf{Model} & \textbf{Protection} & \textbf{PSNR} & \textbf{SSIM} \\ 
        \hline
        \multirow{3}{*}{ip2p}           & w/o Protection      & 4.11          & 0.14          \\
                       & w/ Photoguard       & 4.06          & 0.11          \\
                       & w/ EditShield       & \textbf{3.32}          & \textbf{0.09}          \\ 
        \hline
       \multirow{3}{*}{ip2p-mb}        & w/o Protection      & 19.11         & 0.78          \\
                       & w/ Photoguard       & 17.56         & 0.67          \\
                       & w/ EditShield       & \textbf{16.82}         & \textbf{0.66 }         \\ 
        \bottomrule
    \end{tabular}   } 
    \captionof{table}{Image quality comparison.}
    \label{tab_quality}
\end{minipage}
\end{figure*}

\begin{figure*}[t]
\centering
\begin{minipage}{0.495\linewidth}
\centering
\includegraphics[width=\linewidth]{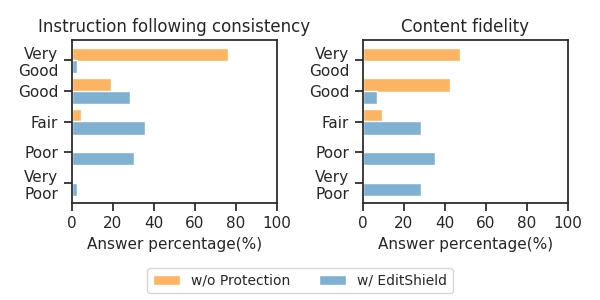}
\caption{Editing evaluation by GPT-4V.}
\label{gpt4v}
\end{minipage}%
\hfill
\begin{minipage}{0.495\linewidth}
\centering
\includegraphics[width=\linewidth]{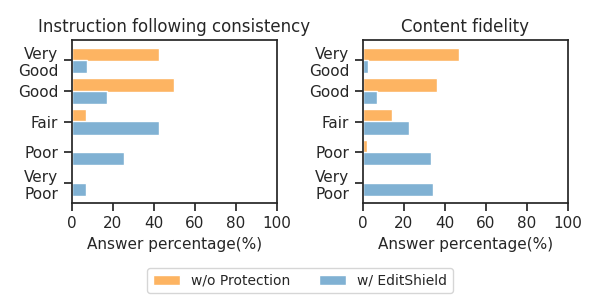}
\caption{Editing evaluation by humans.}
\label{user}
\end{minipage}
\end{figure*}

\subsection{Qualitative Results on Protection}
We provide a qualitative analysis of various images in Fig.~\ref{VIS1}. The source images and their corresponding instructions in the uppermost four rows are from IPr2Pr, while the last two are from MagicBrush. We compare the quality of editing on images without the protection, and with the protection of \method. It can be observed that images derived from protected images exhibit mismatched subjects and low quality compared with those edited from the original source images. 
For instance, in the first row, under the instruction \textit{``turn into a cartoon''}, the girl in the protected image disappears, replaced by a group of cartoon characters. In the last row, the background of the edited image is heavily distorted with blue noise. These results demonstrate that an unauthorized editor is unable to achieve the desired output, thus underscoring the effectiveness of our protection.

\subsection{Quantitative Results on Protection}
\noindent \textbf{Protection Effectiveness.} We calculate two metrics to assess image protection: the cosine similarity of CLIP image embeddings and the directional CLIP similarity, both applied to unprotected and protected images by PhotoGuard and \method under the same instruction. Results are shown in box plots in Fig.~\ref{tradeoff}, where orange, green and blue boxes represent similarities without protection, with PhotoGuard and \method, respectively. The box shows the quartiles of the data while the whiskers extend to show the rest of the distribution, except for points that are ``outliers”.

As indicated by Fig.~\ref{tradeoff}, \method performs better than PhotoGuard, as evidenced by the smaller values in the blue boxes across both similarity metrics. When applying \method, the image editing is less successful. In other words, changes made during the editing of the protected image show less agreement with the given instruction. Furthermore, the protected image after editing exhibits more differences in image content compared with those without protection. This is because \method can shift the latent representation of the original image, resulting in lower image consistency after editing.  

\noindent \textbf{Image Quality Assessment.} To evaluate image quality, we compare the editing results without protection, and with protection by PhotoGuard and \texttt{EditShield}, including Peak Signal-to-Noise Ratio (PSNR) and Structural Similarity Index Measure (SSIM)~\cite{wang2004image}. For both metrics, lower scores indicate worse image quality after editing. Results are presented in Table~\ref{tab_quality}.
As observed, with the protection of \method, edited images consistently show smaller quality scores. Compared with PhotoGuard, \method can result in poorer image quality.

\noindent \textbf{Evaluation by GPT-4V.} We further leverage GPT-4V~\cite{openai2023gpt4v} to comprehensively evaluate the instruction-guided image editing with the protection of \method. It is asked to give quantitative scores on instruction-following consistency and content fidelity. The former represents how much the edited image faithfully follows the given instruction, while the latter evaluates the consistency in subject, style, content, and image quality between the source and the edited image. Both scores are measured on a scale from 0 to 1. For a more specific analysis, we further categorize these scores using a 5-level Likert scale, representing ``very poor'', ``poor'', ``fair'', ``good'' and ``very good''. 

We randomly choose 100 ``source-edited'' image pairs with \method and report evaluation results in Fig.~\ref{gpt4v}. As observed, when \method is applied, GPT-4V's scores drop to a great extent, with a notable increase in scores categorized as ``very poor'' and ``poor'' based on content fidelity and instruction-following consistency. These findings align with our quantitative assessment, underscoring \method's success in preventing unauthorized image modifications by compromising the consistency and fidelity of the edited image. As a result, images edited with the protection of \method exhibit mismatched subjects or unrealistic contents, effectively preventing unauthorized editors from achieving their intended modifications. We will provide details of querying GPT-4V in the Supplementary materials.


\noindent \textbf{Evaluation by Human Users.} We also conduct a human evaluation for comparative analysis. Twenty randomly selected ``source-edited'' image pairs, along with corresponding instructions, are presented to 50 participants. Each participant is tasked with assessing whether the edited image meets the editing requirements, considering content fidelity, instruction following, and image quality. Quantitative scores between 0 and 1 will be given by those participants. We calculate the percentage of five responses and report results in Fig.~\ref{user}. We observe that most images, with the protection of \method, don't align with the editing requirements. In other words, \method effectively prevents editors from achieving their intended editing goals. This outcome not only corroborates the quantitative efficacy of our protection as previously demonstrated but also highlights its practical effectiveness in preserving images against unauthorized modifications. Further details are in Supplementary materials.

\begin{figure*}[t]
\centering
    \subfloat{
        \includegraphics[width=0.49\linewidth]{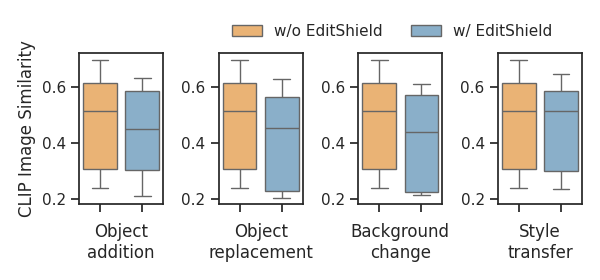}}
    \subfloat{
        \includegraphics[width=0.49\linewidth]{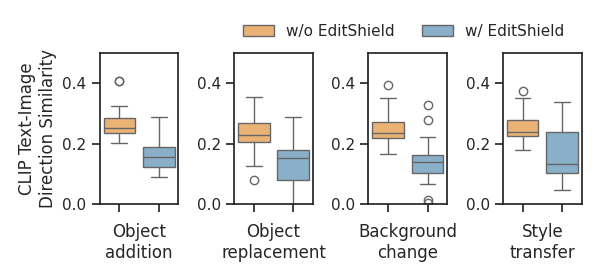}}
\caption{The impact of different types of instructions on \method.}
\label{impact_type}
\end{figure*}

\begin{figure}[t]
\centering
    \subfloat[Editing types]{
        \includegraphics[width=0.49\linewidth]{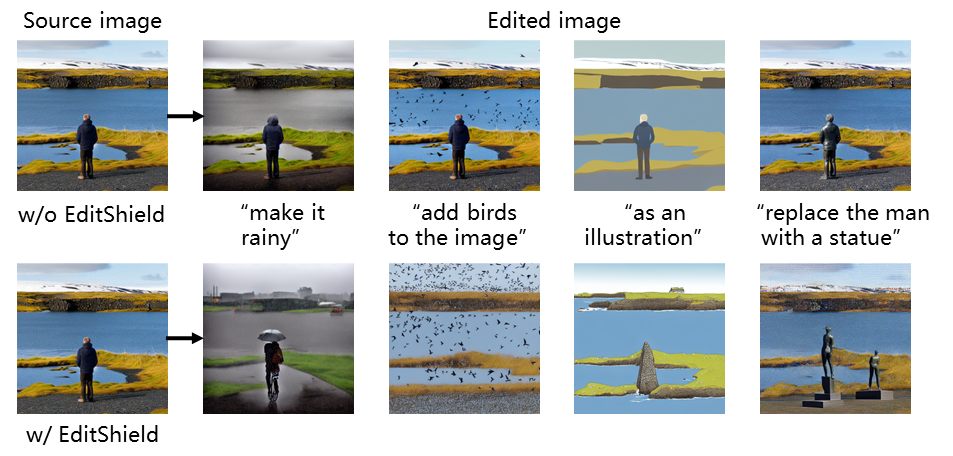}\label{type}}
    \subfloat[Synonymous instruction phrases]{
        \includegraphics[width=0.49\linewidth]{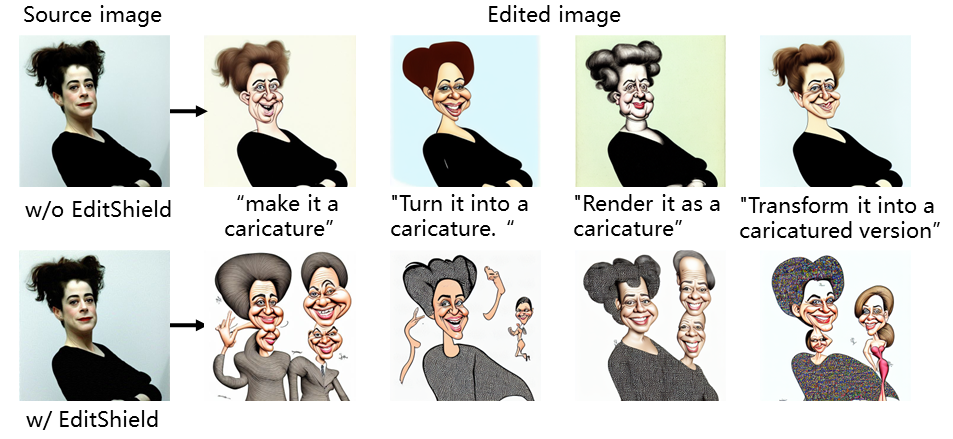}\label{variety}}
\caption{Edited images with \method on different instructions.}
\label{robustness}
\end{figure}

\subsection{Protection Robustness}
\begin{wrapfigure}{r}{0.5\linewidth}
\centering
    \includegraphics[width=1\linewidth]{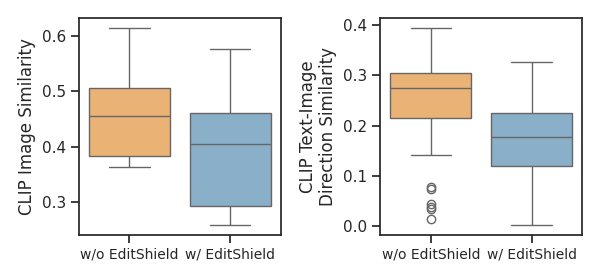}
    \caption{The impact of synonymous instruction phrases on \method.}
    \label{fig:synonym_impact}
\end{wrapfigure}
We verify the protection robustness of \method under different instruction prompts, including different editing types and synonymous instruction phrases.

\noindent \textbf{Editing Types.} We further investigate how the editing instruction types will affect \method. According to~\cite{zhang2023magicbrush}, we focus on four types of instructions: object addition, object replacement, background change and style transfer. We randomly chose 100 images and conducted different types of image editing on them. Two similarity metrics are reported in Fig.~\ref{impact_type}. Besides, we also present some visual analysis in Fig.~\ref{type}. We can observe \method shows robust protection performance against a variety of instruction types, as evidenced by all lower values in blue boxes across both similarity metrics. Notably, the protection is more effective for instructions involving object changes, such as object addition and replacement. This can be attributed to \method's ability to induce mismatches in subjects.

\noindent \textbf{Synonymous Instruction Phrases.} We consider how \method's performance might vary with synonymous instruction phrases, such as using \textit{``make it burn''} instead of \textit{``make it on fire''}. We leverage ChatGPT to generate four synonymous phrases for each original instruction prompt. Then we perform image editing on protected images for comparison. For measurement, we calculate image and direction similarity on average, then present quantitative results and visual examples in Fig.~\ref{robustness} and Fig.~\ref{fig:synonym_impact}, respectively. We observe that the values in the blue boxes remain lower than those in the orange boxes. This suggests that using synonyms in instruction expressions will not affect the effectiveness of \method's protection. 

\subsection{Possible Countermeasures} 
We consider potential countermeasures the unauthorized editor might employ to reduce our protection effectiveness. 

\begin{figure}[t]
\centering
    \subfloat[Spatial smoothing]{
        \includegraphics[width=0.49\linewidth]{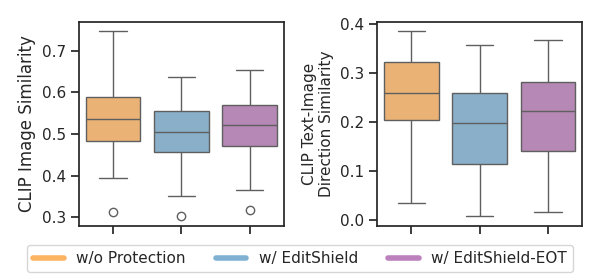}\label{ss}}
    \subfloat[JPEG compression]{
        \includegraphics[width=0.49\linewidth]{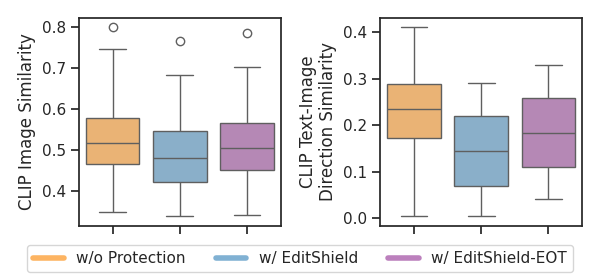}\label{jc}}
\caption{Quantitative results on possible countermeasures.}
\label{adaptive}
\end{figure}

One potential approach involves leveraging image processing methods for protection mitigation, which does not require retraining the generative model or modifying the model structure. After applying \method, spatial smoothing (SS)~\cite{Xu0Q18} and JPEG compression (JC)~\cite{dziugaite2016study} may be adopted by editors to mitigate the protection. Here we set window $size=2$ in SS and $quality=80$ in JC. We calculate two similarity metrics for editing evaluation. Besides, we also remove EOT in \method for comparison. Results are reported in Fig.~\ref{adaptive}. As observed, two metrics drop after our protection, i.e., \method still shows effectiveness in image protection when facing these image transformations. When we remove EOT, protection effectiveness decreases, as purple boxes exhibit larger metrics than blue ones.  
Such observations demonstrate that the integration of EOT can improve \method's protection against image transformations.

\subsection{Ablation Studies}
\begin{figure*}[t]
\centering
\begin{minipage}[c]{0.32\linewidth}
\centering
\includegraphics[width=\linewidth]{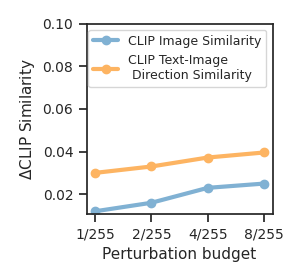}
\caption{Impact of overall perturbation budget.}
\label{budget}
\end{minipage}%
\hfill
\begin{minipage}[c]{0.66\linewidth}
\centering
\includegraphics[width=\linewidth]{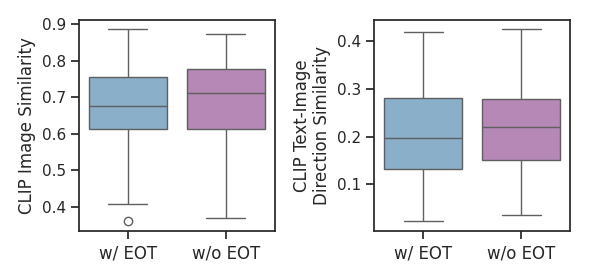}
\caption{Impact of EOT.}
\label{EOT}
\end{minipage}
\end{figure*}

\noindent \textbf{Perturbation Budget.} We investigate the effect of the overall perturbation budget on our protection. Utilizing the IPr2Pr dataset, we explore how varying budgets from 1/255 to 8/255 influence the protection's effectiveness, as measured by the median drop in two critical similarity metrics. Results are shown in Fig.~\ref{budget}. We observe that a relatively modest perturbation budget of 4/255 is sufficient for \method to significantly prevent the editing performance of instruction-guided image editing models, thereby leading to robust protection against unauthorized modifications. 

\noindent \textbf{Impact of EOT}. To evaluate the individual contributions of EOT to its overall effectiveness, we conducted ablation studies using the IPr2Pr dataset. The results, visualized in Fig.~\ref{EOT}, show blue boxes are slightly lower than purple ones in two metrics. This indicates that the integration of subtle image transformations in \method contributes to more effective protection. 


\section{Conclusion}
In this paper, we propose the first protection method against unauthorized editing by instruction-guided diffusion models. The proposed protection \method, crafts small perturbations that try to shift the latent distribution from the source image. When the protection is applied, the edited image will have mismatched subjects and be of low quality. Extensive evaluations of two datasets have verified that \method can effectively prevent images from being edited by instruction-guided diffusion models. Besides, \method is robust towards different editing types and synonymous instruction phrases. 
However, \method shows limitations in protecting images in a black-box manner against online instruction-guided image editing platforms. An important future direction is to design a more efficient and robust perturbation in practice.

\newpage

\section*{Acknowledgements}
We extend our gratitude to all authors, reviewers, and the chair for their invaluable contributions. Additionally, we would like to express our appreciation to Dongping Chen for providing computational resources.

\bibliographystyle{splncs04}
\bibliography{egbib}

\begin{thebibliography}{10}
\bibitem{totonews}
Jiang, K.: These ai images look just like me. what does that mean for the future of deepfakes? Toronto Star
\end{thebibliography}

\end{document}